\documentclass{iopjournal}

\usepackage{array}
\usepackage{lineno}
\usepackage{mathtools}
\usepackage{amssymb}
\usepackage[normalem]{ulem}
\usepackage{xcolor}

\begin{document}

\newcommand{\xf}{x^\prime}

\articletype{Paper} 

\title{General diffraction properties of aperiodic slit arrays}

\author{Thiago S. Ferreira$^{1}$\orcid{0009-0003-3106-1930}, Daniel Jonathan$^1$\orcid{0000-0002-5179-3755}, Antonio Z. Khoury$^{1}$\orcid{0000-0002-7487-5067} and Daniel S. Tasca$^{1,*}$\orcid{0000-0003-1980-2276}}

\affil{$^1$Instituto de Física, Universidade Federal Fluminense, Niterói, Brazil}

\affil{$^*$Author to whom any correspondence should be addressed.}

\email{danieltasca@id.uff.br}

\keywords{diffraction gratings, Fraunhofer diffraction, Fourier transform, aperiodic media, structured beams}

\begin{abstract}
Fraunhofer diffraction plays a vital role in experimental physics not only because it accurately describes the behaviour of light in the usual propagation limit, but also because it links the diffracted light with the scattering object through one of the most important mathematical transformations in physics: the Fourier transform. Acting as a probe in material characterisation as well as used as a tool for particle trapping or sensing, the pattern of interference maxima resulting from the Fraunhofer diffraction through periodic scattering is an ubiquitous routine. In this paper we analyse the Fraunhofer diffraction resulting from the much less studied aperiodic scatter of the light. We provide general conditions for the experimental observation of the peaks of interference maxima featured into patterns that display periodic structures on a number of distance scales. Our theoretical analysis is supported by thorough experimental demonstrations.
\end{abstract}

\section{Introduction}

Light diffraction by periodic media has been studied for almost two and a half centuries. In 1785, F. Hopkinson observed the interference pattern produced by a lamp's light when it passed through a silk handkerchief. The phenomenon was later studied by his friend, the astronomer D. Rittenhouse, who then created his own diffraction grating \cite{Rittenhouse:1786}. Periodic gratings are well understood and widely utilised in various technologies, including laser pulse compression, optical material structuring, and laser linewidth control \cite{Bonod:2016}. In particular, gratings consisting of equally spaced apertures have interesting applications in areas such as spectrometry \cite{Kong:01}, optical communication \cite{Duarte:2011}, and interferometry \cite{Kawashima:20}.

In contrast, diffraction by aperiodic gratings remains less explored. Prior studies have examined their ability to deflect light \cite{Das:2013}, their potential for measuring the radii of Gaussian beams \cite{Mata-Mendez:08}, the focusing properties of chirped gratings \cite{Sanchez-Brea:16}, the potential advantages of Fibonacci gratings in image-forming devices \cite{Verma:2014}, as well as the self-imaging capacities of fractal \cite{Teng:2014,Wang:14} and quasiperiodic \cite{Zhang:13} gratings. Aperiodic gratings have also been employed in the generation of achromatic \cite{Yuan:17} and single-photon \cite{Yuan:16} super-oscillations. X-ray diffraction by aperiodic solids such as quasicrystals, which present long-range order without translational symmetry, has also been a subject of considerable interest in crystallography \cite{Grimm:2015,Strzalka:2016}. For instance, previous studies have investigated wave dynamics in photonic quasicrystals \cite{Freedman:2006}, the optical properties of deterministic aperiodic nanostructures \cite{DalN:2012}, and light transmission through self-similar aperiodic structures \cite{Zotov:2010}.

In Ref. \cite{Maia:22}, we investigated the correspondence between the far-field diffraction pattern of certain aperiodic gratings and the revival phenomena in quantum systems. The properties of these gratings motivated the present work, in which we conduct a more thorough investigation of the far-field diffraction behaviour of a broader set of aperiodic slit gratings. Under certain conditions, their Fraunhofer intensity patterns exhibit quasiperiodic oscillations at different distance scales. We examine these conditions and investigate the possibility of suppressing the side maxima with the sinc function that arises from the Fourier transform of individual slits. Interestingly, we find that achieving such suppression would require part of the slit array to merge into a single aperture. Our theoretical predictions are supported by experimental results. 

Our paper is organised as follows. In section \ref{Sec2} we analyse the Fraunhofer diffraction structure from a general aperiodic slit array. In section \ref{sec:Theory} we provide mathematical conditions for the observation of periodic peaks in the Fraunhofer diffraction pattern conditioned upon the geometrical properties of the arrays and its illumination pattern. Our experimental demonstrations are presented in section \ref{sec:Exp}. Conclusions and discussions are included in section \ref{sec:Conc}.


\section{Fraunhofer diffraction structure of an optical field}
\label{Sec2}
\subsection{The near-field}

A coherent optical field incident upon a flat structure with multiple apertures disposed along a single dimension, $x$, transverse to the beam propagation axis, may be represented as
\begin{equation} \label{near-field}
\Psi(x) \propto \sum_n \psi_n(x), 
\end{equation}
where the field behind each aperture is $\psi_n(x)=A_n(x)e^{i\phi_n(x)}$ and $n\in\mathbb{Z}$ serves as an aperture label. The field modulations considered in this paper are those introduced by apertures of a general, \textit{aperiodic} diffraction grating -- or, in other words, aperiodic slit arrays. Considering a flat-wave front illumination and diffraction gratings with amplitude-only modulations, we shall rewrite $\psi_n(x)$ as
\begin{equation} \label{FieldBehindAperture}
\psi_n(x)=A_n e^{i\phi}\Pi_n(x), 
\end{equation}
where $A_n^2$ may be understood as the transmittance function of slits modeled by rectangular apertures,
\begin{equation} \label{Slit} 
\Pi(x )=\left\{ \begin{array}{ccc}   1, & \,   |x|   \leq   s/2 \\   0, &  {\rm otherwise}  \end{array} \right. .
\end{equation} 
In Eq. \eqref{FieldBehindAperture}, the phase factor $\phi_n(x)=\phi$ was made constant and $\Pi_n(x) \coloneqq \Pi(x-x_n)$ represents a slit of width `$s$' centred at the transverse coordinate $x_n$. Without loss of generality, we take $\phi  = 0$ from this point on. 

\subsection{The far-field}

In the Fraunhofer limit, the interference generated by the field scattered by the multiple apertures given in Eq. \eqref{near-field} is obtained by means of a Fourier transform. Using a Fourier transform lens system, the input (near-field)  and output (far-field) structures relate as follows  
\begin{equation}\label{OpticalFourier}
 \tilde{\Psi}(\xf) =\sqrt{ \frac{\xi}{2\pi }} \int dx \, \Psi(x) e^{-i x (\xi \xf)},
\end{equation}
where $\xi=2\pi/f\lambda$ is a constant with dimension of length$^{-2}$, $f$ is the focal length of the lens and $\lambda$ is the wavelength of the monochromatic light field. In Eq. \eqref{OpticalFourier}, $\xf$ represents the transverse coordinate of the Fourier-transformed field in the back focal plane of the lens.
Inserting Eq. \eqref{near-field} into Eq. \eqref{OpticalFourier} and making use of Eqs. \eqref{FieldBehindAperture} and \eqref{Slit}, we get
\begin{equation}\label{Far-field}
 \tilde{\Psi}(\xf) = \tilde{\Pi}(\xf) \sum_n A_n e^{-i \kappa_n \xf},
\end{equation}
where
\begin{equation} 
\label{kappan} 
\kappa_n \coloneqq \xi x_n =\frac{2\pi}{\lambda f} x_n,
\end{equation} 
and
\begin{equation} 
\label{SlitFourier} 
\tilde{\Pi}(\xf)= s \sqrt{ \frac{\xi}{2\pi }}   \mathrm{sinc} \left( \frac{\xi s \xf }{2} \right),
 \end{equation} 
is the Fourier transform of the rectangular aperture defined in Eq. \eqref{Slit}.

Eq. \eqref{Far-field} describes the Fraunhofer diffraction pattern of a general slit array. For instance, using $x_n=nL$, Eq. \eqref{Far-field} describes the interference pattern from the usual \textit{periodic} diffraction grating with slits spaced by $L$. The Fourier transform of the slit aperture, Eq. \eqref{SlitFourier}, behaves as an envelope that modulates the peaks of the diffraction maxima: the well-known diffraction orders. Before analysing particular examples of gratings, in the next section we shall investigate further general properties of Eq. \eqref{Far-field}.
\section{Far-field diffraction periodicities of general slit arrays}
\label{sec:Theory}

The near-field structure given in Eq. \eqref{near-field} is completely specified by the slit distribution of the array, $x_n$, the slit width, $s$, and the set of field amplitudes illuminating each slit of the array, $A_n$. Without specifying a particular geometry for the grating slit distribution, it is possible to derive some general properties of its Fraunhofer diffraction pattern by imposing only a few natural constraints on the illumination distribution $A_n$.

In our mathematical definition, the index labeling the slits of the array assumes any integer value running from negative to positive infinity. In the laboratory, however, only a finite number of slits are illuminated and therefore it is natural to assume that the illumination distribution is concentrated around some arbitrary slit of the array. As an example, let us consider a Gaussian amplitude distribution centred around  slit number $\bar{n}$:
\begin{equation}\label{A_nGauss}
A_n\propto e^{-\frac{(n-\bar{n})^2}{4\sigma^2}},
\end{equation}
where $\sigma$ is the standard deviation of the \textit{intensity} distribution $A_n^2$. 

It is important to note that, for aperiodic diffraction arrays, the function $x_n$ is nonlinear in the slit number $n$. Thus, the dependence of the field distribution on the transverse coordinate $x$ in real space and on the slit index $n$ are different in general. The two only coincide in the special case where the diffraction grating is periodic, when $x_n = n L$ for some fixed spacing $L$.

In what follows, it suffices to assume that the slit positions $x_n$ interpolate a well-behaved, smooth function $x(n)$, with $n$ viewed as a continuous parameter, such that $x_n$ can be expanded in a Taylor series around the value $\bar{n}$, where $A_n$ is maximal: 
\begin{eqnarray}  \label{TaylorExpansion}
x_n &\approx& \sum_{j=0}^{\infty} \frac{x_{\bar{n}}^{(j)}}{j!} (n-\bar{n})^j \\ \nonumber
&=& x_{\bar{n}}  +  x_{\bar{n}}^{(1)}(n-\bar{n}) +  \frac{x_{\bar{n}}^{(2)}}{2!}(n-\bar{n})^2 + \cdots,
\end{eqnarray} 
where $x_{\bar{n}}^{(j)}$ denotes the $j$-th derivative of $x(n)$, evaluated at $n = \bar{n}$:
\begin{equation}  \label{Derivatives}
x_{\bar{n}}^{(j)} \coloneqq \left. \frac{d^{(j)} x(n) }{dn^{(j)}} \right|_{n=\bar{n}}.
\end{equation} 
Using the Taylor expansion \eqref{TaylorExpansion} and definition \eqref{Derivatives}, we can rewrite the Fraunhofer diffraction pattern given in Eq. \eqref{Far-field} as
\begin{equation}\label{Far-fieldTaylor}
 \tilde{\Psi}(\xf) = \tilde{\Pi}(\xf)  \exp\left[ -i \frac{2\pi}{L^\prime_0} \xf \right]
 \sum_n A_n \exp\left[ -i \sum_{j=1}^{\infty}\frac{2\pi}{L^\prime_j} (n-\bar{n})^j \xf \right],
\end{equation}
where $L^\prime_{j}$ is a parameter with dimension of length:
\begin{equation}  \label{OscillationPeriods}
L^\prime_{j} \coloneqq  \frac{2\pi }{\xi |x_{\bar{n}}^{(j)}|}j! = \frac{\lambda f}{|x_{\bar{n}}^{(j)}|}j!\;.
\end{equation} 
Whenever $\xf$ takes on a value that is a multiple of $L_j'$, each of the terms in Eq. \eqref{Far-fieldTaylor} with different `$n$', but the same `$j$', interfere constructively. These quantities may therefore be identified as periodicity scales within the far-field diffraction pattern. If, furthermore, these scales differ in order of magnitude, that is, if
\begin{equation} \label{TjRatio}
    \frac{L^\prime_{j+1}}{L^\prime_j} = (j+1) \left| \frac{ x^{(j)} _{\bar{n}} }{ x^{(j+1)} _{\bar{n}} }\right| >> 1\;,
\end{equation}
then we can expect to see distinct maxima in the diffraction pattern, separated by each periodicity.

In many cases, condition \eqref{TjRatio} reduces to a restriction on the value of $\bar{n}$, such as, e.g.,  $\bar{n} \gg 1$ \cite{Maia:22}. In other words, for many discrete distributions $x_n$, condition \eqref{TjRatio} is satisfied by appropriately choosing the value of $\bar{n}$. Thus, the Taylor expansion of the slit distribution, Eq. \eqref{TaylorExpansion}, allows us to identify many different periodicities in the Fraunhofer diffraction pattern of general slit arrays, even when the slit distribution is itself an aperiodic function. We shall discuss a specific example of condition \eqref{TjRatio} in the next section.

It is useful at this point to check the special case of a periodic diffraction grating, with slits separated by a constant spacing $L$. In this case, as is well known, the far-field diffraction pattern is also structured into periodic interference maxima, whose period  corresponds here to the first-order periodicity
\begin{equation} 
\label{Tprime} 
L_1^\prime \coloneqq \frac{2\pi}{\xi L}.
\end{equation}
Meanwhile, in this case the higher-order periodicities diverge ($L^\prime_{j>1} \rightarrow \infty$) and are therefore not observable. 

The intensity of the Fraunhofer diffraction pattern is obtained by taking the squared modulus of the amplitude given in Eq. \eqref{Far-fieldTaylor}:
\begin{equation}\label{FFIntensity}
   I(\xf) =  |\tilde{\Psi}(\xf)|^2 = \tilde{\Pi}^2(\xf) \sum_n \sum_m A_n A_m \exp\left[ -i \sum_{j=1}^{\infty}\frac{2\pi}{L^\prime_j} (n-\bar{n})^j \xf \right]\exp\left[ i \sum_{j=1}^{\infty}\frac{2\pi}{L^\prime_j} (m-\bar{n})^j \xf \right]\;,
\end{equation}
which can be reduced, using the symmetry under the exchange $n \leftrightarrow m$, to: 
\begin{equation}\label{FFIntensity2}
     I(\xf) = \tilde{\Pi}^2(\xf) \sum_n \sum_m A_n A_m \cos{ \Bigg\{ \sum_{j=1}^{\infty}\frac{2\pi}{L^\prime_j}\left[(n-\bar{n})^j- (m-\bar{n})^j \right]\xf  \Bigg\}} \;.
\end{equation}
As in the case of the amplitude distribution, Eq. \eqref{Far-fieldTaylor}, the parameters $L_j^\prime$ with $j \ge 1$ also appear as periodicities in the intensity pattern, Eq. \eqref{FFIntensity2}. By contrast, the $0$-th order periodicity ($L_0^\prime$) is not directly observable in intensity. The experimental observation of this lowest order periodicity is discussed in Subsection \ref{ZerothOrder}.

\subsection{Conditions for experimental observation}
\label{subsecConditions}

As discussed above, the far-field intensity pattern of aperiodic gratings may present diffraction peaks at distance scales that are multiples of $L^\prime_1$, $L^\prime_2$, etc. In Ref. \cite{Maia:22}, we have shown that the diffraction features associated with the first-order periodicity are generally easily observable, whereas the higher-order periodicities may also be observable depending on the particular kind of the aperiodic slit distribution. 

Here we would like to establish a more general mathematical condition for the existence of these features in the diffraction pattern. This condition arises from the enveloping sinc function presented in Eq. \eqref{SlitFourier}, which approaches zero as $|x^\prime|$ increases, limiting the number of observable maxima in the far-field intensity pattern. A given order periodicity could even be entirely suppressed if its peaks coincide with the zeros of the sinc function. Accordingly, as a figure of merit for this analysis, we obtain the physical conditions for which the peaks of interference maxima associated with the many orders of Eq. \eqref{OscillationPeriods} exist within the central region defined by the zeroth diffraction order. Thus, matching $L^\prime_ j$ with the first zero of Eq. \eqref{SlitFourier}, we get
\begin{equation}\label{SupressionCond}
    L^\prime_j =\frac{2 \pi}{\xi s}   \Rightarrow s = \frac{1}{j!} \left| x^{(j)} _{\bar{n}} \right|\;.
\end{equation}
As expected, Eq. \eqref{SupressionCond} sets a condition that relates the width of the slits and the parameters characterising the slit distribution (through the derivatives of $x_n$). When the equality is satisfied, the $j$-th order periodicity completely disappears from the far-field intensity pattern. From Eq.~\eqref{TjRatio}, it is clear that, in terms of experimental constraints, the most restrictive instance of this relation appears for the lowest-order periodicity ($j=1$), since it requires the largest slit width necessary to satisfy the equation. Using $j=1$ in Eq. \eqref{SupressionCond}, we conclude that to observe the first diffraction peak associated with $L^\prime_1$ within the zeroth diffraction order imposed by the slit width, the slit width must be chosen such that $s \lesssim \left| x^{(1)} _{\bar{n}} \right|\ $. 

It is now useful to analyse the implications of this condition for the grating's geometry. For this purpose, let us suppose that we want to design the diffraction grating such that Eq. \eqref{SupressionCond} is exactly satisfied for $j=1$. To better understand this condition, let us first notice that for the case of a periodic grating, the condition implies the use of $s = L$, transforming the slit array into a completely opened aperture. We therefore conclude that, as long as the grating remains a periodic array of separate slits, at least two diffraction peaks would be visible at the transverse positions $\xf = \pm L^\prime_1$ within the zeroth diffraction order imposed by the slit width.

We now investigate whether a similar situation occurs for general aperiodic slits arrays. First, let us determine the conditions under which the central slit for the Gaussian illumination, $\bar{n}$, and its neighbour, $\bar{n} + 1$, are separated by a distance equal to their width, forming a single aperture. Mathematically, this corresponds to: $|x_{\bar{n}+1} - x_{\bar{n}}| = s$. We then expand $x_{\bar{n} + 1}$ in a Taylor series around $\bar{n}$ to rewrite this difference as
\begin{eqnarray}  \label{SlitsSep}
|x_{\bar{n}+1} - x_{\bar{n}}| &=& \left| \sum_{j= 0}^{\infty} \frac{x^{(j)}_{\bar{n}}}{j!} - x_{\bar{n}} \right| \\ \nonumber  &\approx& \left| x^{(1)}_{\bar{n}} + \mathcal{O}\left(x^{(2)}_{\bar{n}} \right) \right| \;.
\end{eqnarray}

In order to better understand the implications of Eq. \eqref{SlitsSep}, let us recall that we assume the validity of the inequality part of \eqref{TjRatio}. For this reason, the right-hand side of \eqref{SlitsSep} can be approximated by $|x^{(1)}_{\bar{n}}|$ which, in turn, is exactly the condition that satisfies Eq. \eqref{SupressionCond} for $j=1$. We thus conclude that as long as the slit $\bar{n}$ (the slit at which illumination is maximal) remains separated from its neighbours, the diffraction peaks at $\xf = \pm L^\prime_1$ will be visible within the zeroth diffraction order.

\subsection{Experimental observation of the $0$-th order periodicity}
\label{ZerothOrder}

The $0$-th order periodicity, which appears as a global phase factor in Eq. \eqref{Far-fieldTaylor}, is not observable through intensity measurements alone. In order to observe diffraction features at this distance scale one can interfere the light going through the array with a mirror copy of itself, as shown in Ref. \cite{Maia:22}. This situation is described by a mirror-symmetric array ($x_{-n} = - x_n$) and an illumination pattern that is centred at both $\bar{n}$ and $-\bar{n}$, so that $A_n = A_{-n}$. 

Let $\frac{1}{2}\Psi_r(x)$ denote the near-field structure on the right-hand side of $x = 0$, so that $\Psi(x) = \frac{1}{2}[\Psi_r(x) + \Psi_r(-x)]$. From Eq. \eqref{Far-fieldTaylor} it is easy to see that $\tilde{\Psi}_r(-\xf) = \tilde{\Psi}_r^*(\xf)$, where the asterisk denotes the complex conjugation operation. As a result, the far-field can be expressed as $\tilde\Psi(\xf) = \frac{1}{2}\left[\tilde{\Psi}_r(\xf) +  \tilde{\Psi}_r^*(\xf)\right] = \Re{[\tilde{\Psi}_r(\xf)]} $, with $\Re$ denoting the real part of the function:
\begin{equation}\label{MSFar-Field}
\tilde{\Psi}(\xf) = \Re{[\tilde{\Psi}_r(\xf)]} = \tilde{\Pi}(\xf)\sum_n A_n \cos{\left[ \sum_{j = 0}^\infty \frac{2\pi}{L^\prime_j} (n - \bar{n})^j \xf \right]}\;,
\end{equation}
with $L_0^\prime$ now appearing in the sum inside the argument of cosine, instead of as a global phase. 

In the next section, we shall use Eq. \eqref{MSFar-Field} to represent the Fraunhofer diffraction pattern of mirror-symmetric fields experimentally generated through aperiodic slit arrays. In order to observe the diffraction features discussed in this work, we consider a slit distribution for which the different-order periodicities differ in scale, that is, satisfying the condition in Eq. \eqref{TjRatio}. Furthermore, the distance between neighbouring slits in the chosen distribution decreases with the slit index, allowing for the suppression of first-order maxima through an appropriate choice of parameters. In particular, the chosen array obeys the following slit distribution
\begin{equation} \label{Lsqrtn}
x_{\pm n}=\pm L \sqrt{n}\;.
\end{equation}
The periodicities of the far-field diffraction pattern can be determined by substituting the distribution given in Eq. \eqref{Lsqrtn} into Eq. \eqref{OscillationPeriods}. The corresponding zeroth-, first-, and second-order periodicities are given by 
\begin{align}
    L^\prime_0 &= \frac{2\pi}{\xi L} \bar{n}^{-\frac{1}{2}} \label{L0}\;, \\ 
    L^\prime_1 &=  \frac{2\pi}{\xi L}2\bar{n}^{\frac{1}{2}} \;,\label{L1}\\
    L^\prime_2 &=  \frac{2\pi}{\xi L}8\bar{n}^{\frac{3}{2}} \label{L2}\;. 
\end{align}
Let us examine the conditions under which these periodicities are well separated in scale. The ratios between the successive order periodicities are $\frac{L^\prime_1}{L^\prime_0} = 2\bar{n}$ and  $\frac{L^\prime_2}{L^\prime_1} = 4\bar{n}$. Thus, the condition in Eq. \eqref{TjRatio} is satisfied when $\bar{n} \gg 1$. 

Finally, it is worth remarking that, for a sufficiently smoothly varying amplitude $A_n$, it is possible to obtain approximate analytical descriptions of the sum in Eq. \eqref{MSFar-Field}, each of which is valid for a separate diffraction peak, by using the Poisson summation formula \cite{Courant:89} and making a stationary-phase approximation. See, e.g., Refs. \cite{Fleischhauer:93}, \cite{Jonathan:99} for details, and Ref. \cite{Maia:22} for an application to the diffraction structure described by Eq. \eqref{Lsqrtn}.

\section{Experimental results}
\label{sec:Exp}

\begin{figure}[htbp]
\centering\includegraphics[width=12cm]{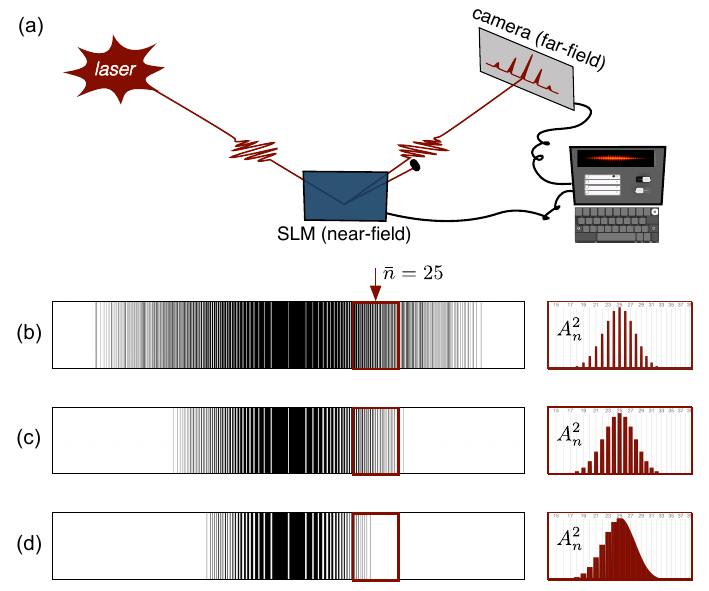}
\caption{(a) Simplified version of our experimental setup. A laser beam ($\lambda = 660$nm) is sent to a Spatial Light Modulator (SLM). The unmodulated light is filtered out, and a Complementary Metal-Oxide-Semiconductor (CMOS) camera records the Fraunhofer diffraction pattern of the modulated light. We also present images of the aperiodic gratings for a slit distribution described by Eq. \eqref{Lsqrtn}, with $L=(0.560 \pm 0.008)$mm and slit widths of (b) $s = 0.024$mm (3 pixels of the SLM); (c) $s = 0.040$mm (5 pixels of the SLM); (d) $s = 0.056$mm (7 pixels of the SLM). The Gaussian illumination in the outlined regions is shown on the right-hand side of the figure for $\bar{n} = 25$, and $\sigma = 2\sqrt{2}$, with the position of the centre of each slit represented by a vertical line. Note that the region of the array where the slits are merged increases as the slits get wider. The parameters used to draw parts (b), (c) and (d) of this figure correspond to the parameters used in the measurements shown in Fig. \ref{FigS}.}
\label{FigureSetup}
\end{figure}

\begin{figure}[htbp]
\centering\includegraphics[width=1\linewidth]{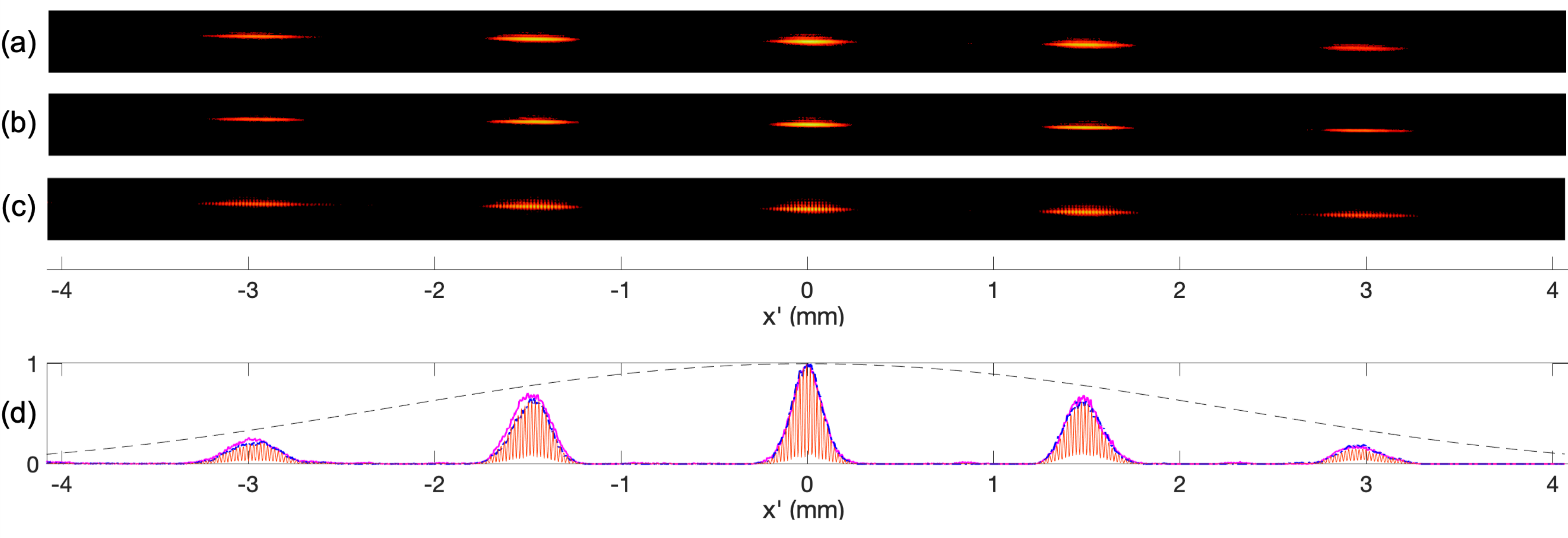}
\caption{Experimental results obtained using a grating with slit distribution given by Eq. \eqref{Lsqrtn} with $L=(0.800 \pm 0.008)$mm and slit width $s=0.024$mm (corresponding to 3 pixels of our SLM). The images show the Fraunhofer diffraction pattern obtained when illuminating the (a) right-hand side ($\bar{n}=20$), (b) left-hand side ($\bar{n}=-20$), and (c) both sides of the grating. A standard deviation of $\sigma = \sqrt{2}$ was used for the Gaussian illumination. In (d) we plot the graphs resulting from the sum of the recorded images over 80 pixels in the vertical direction, each of which normalised to its maximum value. The solid pink, dash-dotted blue, and the thinner solid-red lines correspond to the images (a), (b), and (c) respectively. The black-dashed line corresponds to the plot of the sinc function, Eq. \eqref{SlitFourier}, with the experimental value of the slit width used in this experiment.} 
\label{FigureLR}
\end{figure}

We have experimentally investigated the Fraunhofer diffraction pattern of aperiodic slit arrays. In our setup, a fibre-coupled laser beam with $\lambda=660$nm [Thorlabs S1FC660] is first expanded and sent to a spatial light modulator (SLM). The SLM [HoloEye Pluto NIR-015] is programmed to generate the field given in Eq. \eqref{near-field} that is then Fourier-transformed using a lens system with effective focal length $f=200$mm. The Fraunhofer diffraction pattern is then registered using a Complementary Metal-Oxide-Semiconductor (CMOS) imaging sensor [Basler ace U acA5472-17um]. A sketch of our experimental setup is shown in Fig. \ref{FigureSetup}-(a) (see caption for details).

The chosen slit array has an aperiodic slit distribution described by Eq. \eqref{Lsqrtn} and the amplitude distribution of the light going through the slit array is chosen to be the Gaussian function given in Eq. \eqref{A_nGauss} centred around two opposite values of the slit number: $\bar{n}$ and $-\bar{n}$. Figs. \ref{FigureSetup}-(b)-(d) present examples of gratings with slit positions following Eq. \eqref{Lsqrtn} (on the left), alongside their respective Gaussian intensity distributions (on the right). At first glance, the slit distributions of the illuminated parts may appear periodic, as the aperiodicity becomes more noticeable when analysing a larger portion of the gratings. However, the following experimental results show that the far-field diffraction patterns differ significantly from those of periodic gratings.

In Fig. \ref{FigureLR} we show the Fraunhofer diffraction pattern generated through the illumination of the (a) right-hand side, (b) left-hand side, and (c) both sides of the aperiodic slit array. The identical diffraction patterns registered in the images of Figs. \ref{FigureLR}-(a) and \ref{FigureLR}-(b) reflect the symmetry discussed in Subsection \ref{ZerothOrder}: $\tilde{\Psi}_r(x')$ and its mirror inversion $\tilde{\Psi}_r(-x')$ have the same absolute value.  The rapid oscillations in the diffraction pattern shown in Fig. \ref{FigureLR}-(c) are the result of interference of the fields coming from each side of the array. To better visualise the periodic scales at which the diffraction peaks occur, we plot in Fig. \ref{FigureLR}-(d) the curves corresponding to the sum of the recorded images over $80$ pixels in the vertical direction. The solid pink, dash-dotted blue, and solid red curves correspond to the images (a), (b), and (c), respectively. The Fourier transform of the slit aperture, Eq. \eqref{SlitFourier}, is plotted (the black-dashed line) to serve as a reference, showing that all measured diffraction peaks lie within the zeroth diffraction order. The experimental parameters of the slit array used in these measurements are listed in the caption of Fig. \ref{FigureLR}.


\begin{table}[htbp]
\caption{Far-field periodicities produced by the slit array given in Eq.~\eqref{Lsqrtn}: symbolic expressions presented in Eqs. \eqref{L0}-\eqref{L2}, values calculated from experimental parameters, and values obtained from the fitting of the measurements shown in Fig.~\ref{FigureLR}.}
\label{Tab:Experiment1}
\centering
\renewcommand{\arraystretch}{1.8}
\begin{tabular}{|>{\centering\arraybackslash}p{1.0cm}
                |>{\centering\arraybackslash}p{2.2cm}
                |>{\centering\arraybackslash}p{2.3cm}
                |>{\centering\arraybackslash}p{2.3cm}
                |>{\centering\arraybackslash}p{2.3cm}
                |>{\centering\arraybackslash}p{2.3cm}|}
\hline
 &
 \bfseries Symbolic expressions &
 \bfseries Calculated (mm) &
 \bfseries Fitted Fig.~\ref{FigureLR}-(a) (mm) &
 \bfseries Fitted Fig.~\ref{FigureLR}-(b) (mm) &
 \bfseries Fitted Fig.~\ref{FigureLR}-(c) (mm) \\
\hline\hline

\boldmath $L_0^\prime$ & $\frac{2\pi}{\xi L} \bar{n}^{-\frac{1}{2}}$ &
$0.0369 \pm 0.0007$ &
-- &
-- &
$0.0370 \pm 0.0007$ \\

\hline

\boldmath $L_1^\prime$ & $\frac{2\pi}{\xi L}2\bar{n}^{\frac{1}{2}}$ &
$1.48 \pm 0.01$ &
$1.49 \pm 0.01$ &
$1.48 \pm 0.01$ &
$1.47 \pm 0.01$ \\

\hline

\boldmath $L_2^\prime$ & $\frac{2\pi}{\xi L}8\bar{n}^{\frac{3}{2}}$ &
$118 \pm 1$ &
-- &
-- &
-- \\

\hline
\end{tabular}
\end{table}


The diffraction patterns shown in Fig. \ref{FigureLR} present peaks of interference maxima disposed at two different distance scales. These distance scales correspond to the zeroth-order periodicity, $L_0^\prime$, and the first-order periodicity, $L_1^\prime$, which are discussed in Section \ref{sec:Theory}. The symbolic expressions for the far-field periodicities associated with the slit distribution used in our experiment, given in Eqs. \eqref{L0}-\eqref{L2}, allow us to compute the theoretical values of these
 distance scales by substituting the parameters used in the experiment. These values are listed in Table \ref{Tab:Experiment1}. To estimate the corresponding values from the experimentally measured diffraction patterns, we fit the curves shown in Fig. \ref{FigureLR}-(d) using the absolute square of the Fraunhofer-diffracted field given in Eq. \eqref{Far-field} with the sum in $n$ truncated at a sufficiently large value. The free parameters used in the fitting were $\bar{n}$ (a dimensionless parameter) and $L$ (a parameter with dimension of length), from which the length scales $L_j^\prime$ written in Table \ref{Tab:Experiment1} were calculated. Using the values of $L_j^\prime$ listed in Table \ref{Tab:Experiment1}, we conclude that for this experiment we get $L_0^\prime \approx 37 \mu$m, $L_1^\prime= 40L_0^\prime \approx1.48$mm and $L_2^\prime \approx 80L_1^\prime \approx 1.18$m. It should be noted that although the chosen slit array geometry predicts a finite value for $L_2^\prime$, this length scale is not experimentally observable.

Device resolution, such as the pixel size of modulators and cameras, sets the utmost limit when structuring or measuring light fields. For example, in our experimental setup, the distance between any pair of slits is resolution-limited by the pixel size of our SLM (we use the pixel size as the uncertainty for the length parameter $L$ of the slit array produced). Unlike this \textit{experimental} limitation, in Subsection \ref{subsecConditions} we have shown that designing an arbitrary aperiodic array with geometry satisfying Eq. \eqref{SupressionCond} for $j=1$ leads to a \textit{fundamental} resolution limitation. The suppression of the maxima associated with the first-order periodicity by the zeros of the sinc function requires the width of the slits to be greater than the distance between some of the illuminated slits in the array. This condition implies that the part of the array with merged slits loses the aperiodic structure that is responsible for the Fraunhofer diffraction patterns studied in this paper. The resulting aperture is then described by a set of separate slits distributed aperiodically and a large aperture produced by a number of merged slits, as illustrated in Fig. \ref{FigureSetup}-(d).

\begin{figure}[t]
\centering\includegraphics[width=1\linewidth]{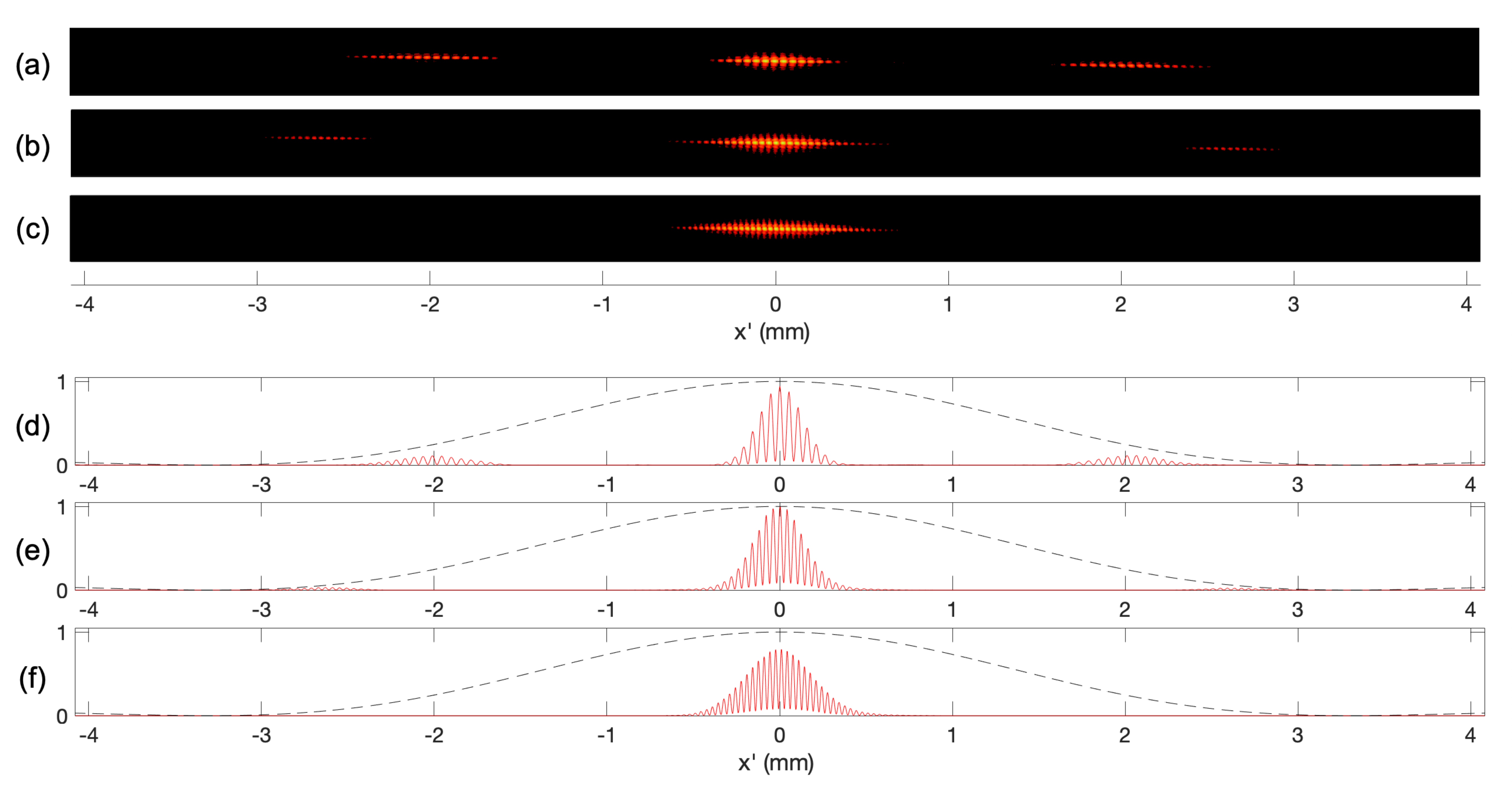}
\caption{Experimental results obtained using a grating with slit distribution given by Eq. \eqref{Lsqrtn} with $(L=0.400 \pm 0.008)$mm and slit width $s=0.04$mm (corresponding to 5 pixels of our SLM). The images correspond to the Fraunhofer diffraction patterns obtained by adjusting the illumination coefficients $A_n$ as the Gaussian function, Eq. \eqref{A_nGauss}, with $\sigma = \sqrt{2}$ and (a) $\bar{n}=\pm 10$, for (b) $\bar{n}=\pm 17$, and (c) $\bar{n}=\pm 25$. In (d), (e) and (f), we plot the graphs resulting from the sum of recorded images (a), (b) and (c), respectively, over 80 pixels in the vertical direction. Each graph was normalised so that the area under the curve equals one and then divided by the maximum value among all three graphs. The black-dashed line corresponds to the plot of the sinc function described by Eq. \eqref{SlitFourier}}. 
\label{FIgN}
\end{figure}

Recalling that condition \eqref{SupressionCond} was established by matching $L_j^\prime$ with the zeros of the diffraction order envelope, let us apply it using $j=1$ to the slit array used in our experiments, Eq. \eqref{Lsqrtn}:
\begin{equation}\label{LsqrtnSuppression}
 L_1^\prime = \frac{2\pi}{\xi L} 2 \sqrt{\bar{n}} \leqslant \frac{2\pi}{\xi s}  \Rightarrow    s \leqslant \frac{L}{2\sqrt{\bar{n}}},
\end{equation}
where we use the inequality symbol to write the geometry condition to observe $L_1^\prime$ within the zeroth order of the diffraction envelope. In the following experimental results, we measure the behaviour of the Fraunhofer diffraction pattern as the grating parameters approach the upper limit of inequality \eqref{LsqrtnSuppression}. For a fixed slit width $s$, we can approach this limit by decreasing the length parameter $L$ or increasing $\bar{n}$. In both of these cases, the zeroth diffraction order remains fixed as we bring $L_1^\prime$ closer to its first zeros. In Fig. \ref{FIgN} we show experimental results for a slit array with $s=0.04$mm, $(L=0.400 \pm 0.008)$mm and $\bar{n}=\pm 10$ [Fig. \ref{FIgN}-(a)], $\bar{n}=\pm 17$ [Fig. \ref{FIgN}-(b)], and $\bar{n}=\pm 25$ [Fig. \ref{FIgN}-(c)]. The corresponding curves obtained by summing over the vertical direction are shown in Figs. \ref{FIgN}-(d), -(e) and -(f). The diffraction pattern shown in Fig. \ref{FIgN}-(c) is obtained from a slit array with condition \eqref{LsqrtnSuppression}  satisfied at its upper limit, $s=L/2\sqrt{\bar{n}}$. We thus see that in this case the peaks of diffraction maxima corresponding to $L_1^\prime$ are completely suppressed by the zeros of the sinc function. It is also visible that, as the side peaks are suppressed, the field's energy concentrates at the central peak, leading to its widening as $\bar{n}$ approaches $\pm 25$.

\begin{figure}[t]
\centering\includegraphics[width=1\linewidth]{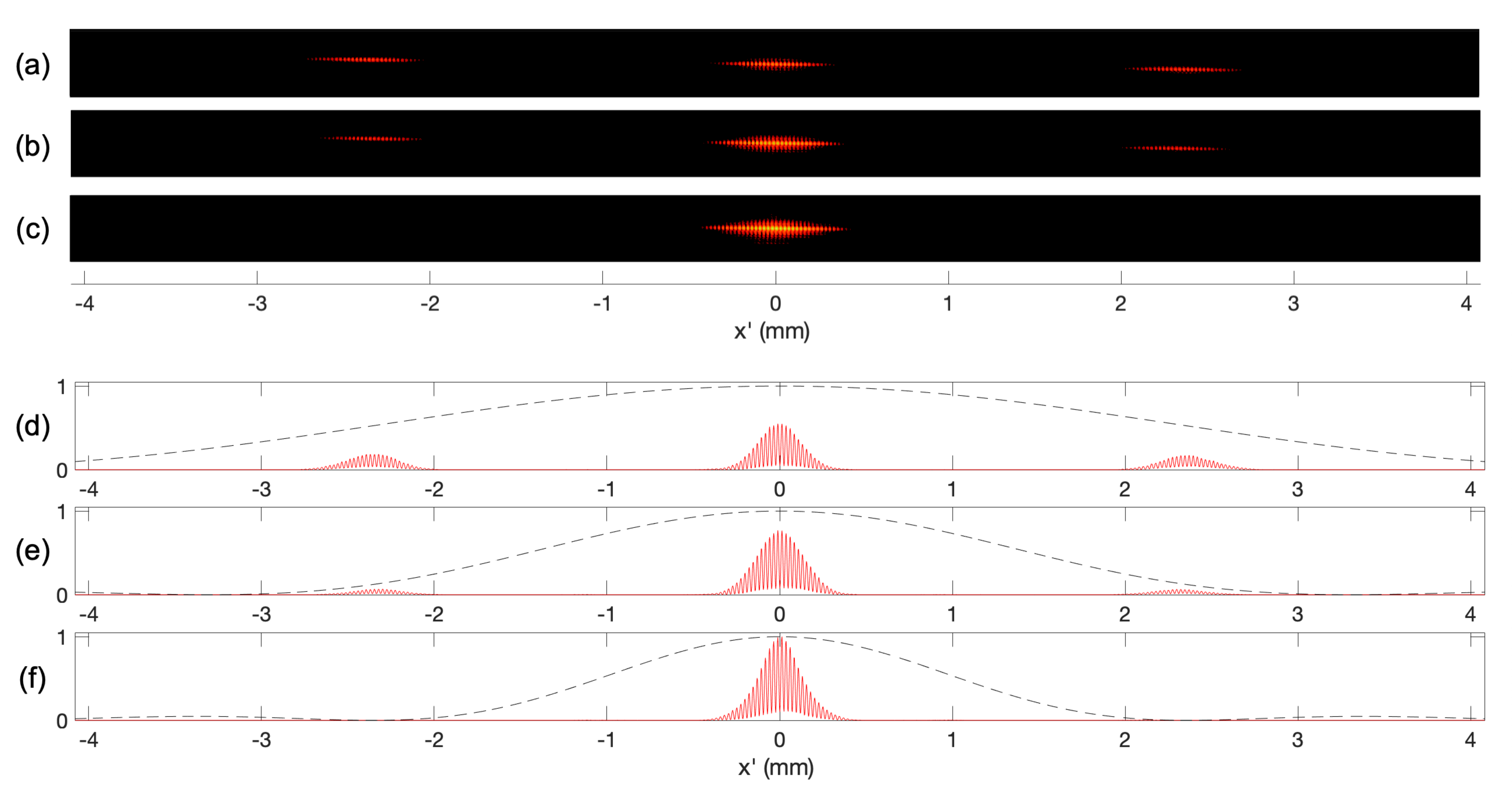}
\caption{Experimental results obtained using a grating with slit distribution given by Eq. \eqref{Lsqrtn} with $L=(0.560 \pm 0.008)$mm, a Gaussian illumination distribution with $\bar{n}=\pm 25$ and $\sigma = 2\sqrt{2}$ The images correspond to the Fraunhofer diffraction patterns obtained with three different slits widths: (a) $s = 0.024$mm (3 pixels of the SLM); (b) $s = 0.040$mm (5 pixels of the SLM); (c) $s = 0.056$mm (7 pixels of the SLM). In (d), (e) and (f), we plot the graphs resulting from the sum of recorded images (a), (b) and (c), respectively, over 80 pixels in the vertical direction. Each graph was normalised so that the area under the curve equals one and then divided by the maximum value among all three graphs. The dashed black line corresponds to the plot of the sinc function described by Eq. \eqref{SlitFourier}. The corresponding aperiodic gratings used in this experiment are illustrated in Figs. \ref{FigureSetup}-(b), (c) and (d).}
\label{FigS}
\end{figure}

Another way to approach the upper limit of \eqref{LsqrtnSuppression} is by increasing the width of the slit $s$ while maintaining fixed values for both $L$ and $\bar{n}$. In this case, $L_1^\prime$ remains constant, while the positions of the zeros of the sinc function decrease. In Fig. \ref{FigS} we present the experimental results for a grating with $L=(0.560 \pm 0.008)$mm, $\bar{n} = \pm 25$ and $s = 0.024$mm [Fig. \ref{FigS}-(a)], $s = 0.040$mm [Fig. \ref{FigS}-(b)] and $s = 0.056$mm [Fig. \ref{FigS}-(c)], with the third case corresponding to total suppression of the diffraction maxima at $\xf = \pm L_1^\prime$. The respective sums over the vertical direction are shown in  Figs. \ref{FigS}-(d), -(e) and -(f). In this case, the reduction of the lateral maxima causes the central peak to become visibly taller. The corresponding aperiodic gratings used in this experiment are illustrated in Figs. \ref{FigureSetup}-(b), (c) and (d).

\section{Conclusions}
\label{sec:Conc}

We have theoretically and experimentally investigated the Fraunhofer diffraction patterns produced by aperiodic slit arrays. We provide general conditions for the observation of quasiperiodic oscillations over different length scales based on the geometry and illumination pattern of the array. Most interestingly, aperiodic diffracting structures with monotonically decreasing distance between adjacent slits suffer from a \textit{fundamental} resolution issue: there comes a point in which the separation of the slits must be smaller than the width of the slits themselves. In this work, we have shown that adjustment of the grating geometry to suppress the peaks of the diffraction maxima associated with its aperiodicity necessarily requires some of the slits to merge into a completely opened aperture. These results provide new insights into the far-field diffraction properties of aperiodic gratings, with possible applications in beam shaping, the design of diffractive devices, and the characterisation of materials with quasiperiodic structures. Finally, we note that in this work we have not discussed the near-field diffraction patterns generated by these aperiodic structures, which feature distorted Talbot patterns that may be of independent interest (for an example, see Ref. \cite{Maia:22}, Fig. 3-(c)). We leave this discussion for future work.

\funding{Fundação Carlos Chagas Filho de Amparo à Pesquisa do Estado do Rio de Janeiro (E-26/201.414/2021); Conselho Nacional de Desenvolvimento Científico e Tecnológico (312478/2023-2); Instituto Nacional de Ciência e Tecnologia de Informação Quântica (465469/2014-0).}

\data{The data supporting this study are available in the Zenodo repository at DOI: 10.5281/zenodo.18476750.}

\providecommand{\newblock}{}

\end{document}